\documentclass{iau307}
\usepackage{graphicx}
\usepackage{natbib}
\usepackage{url}
\usepackage{dtklogos}
\usepackage{amsmath}
\usepackage{amssymb}
\bibpunct{(}{)}{;}{a}{}{,}

\title[1D - 3D] %% give here short title %%
{Linking 1D Stellar Evolution to 3D Hydrodynamic Simulations}

\author[A. Cristini, R. Hirschi, C. Georgy, C. Meakin, D. Arnett, M. Viallet]   %% give here short author list %%
{A. Cristini$^1$, R. Hirschi$^{1,2}$, C. Georgy$^1$, C. Meakin$^3$, D. Arnett$^3$
 \and M. Viallet$^4$}

\affiliation{$^1$Astrophysics group, Keele University, Lennard-Jones Building, Keele, ST5 5BG, UK \\ email: {\tt a.j.cristini@keele.ac.uk} \\[\affilskip]
$^2$Kavli IPMU (WPI), The University of Tokyo, Kashiwa, Chiba 277-8583, Japan \\[\affilskip]
$^3$Department of Astronomy, University of Arizona, Tucson, AZ 85721, USA \\[\affilskip]
$^4$Max-Planck-Institut f\"{u}r Astrophysik, Garching, D-85741, Germany}

\pubyear{2014}
\volume{307} 
\pagerange{}
\jname{New windows on massive stars: asteroseismology, interferometry, and spectropolarimetry}
\editors{G. Meynet, C. Georgy, J.H. Groh \& Ph. Stee, eds.}
\begin{document}

\maketitle

\begin{abstract}
In this contribution we present initial results of a study on convective boundary mixing (CBM) in massive stellar models using the GENEVA stellar evolution code \citep{2008Ap&SS.316...43E}. Before undertaking costly 3D hydrodynamic simulations, it is important to study the general properties of convective boundaries, such as the: composition jump; pressure gradient; and `stiffness'. Models for a 15M$_\odot$ star were computed. We found that for convective shells above the core, the lower (in radius or mass) boundaries are `stiffer' according to the bulk Richardson number than the relative upper (Schwarzschild) boundaries. Thus, we expect reduced CBM at the lower boundaries in comparison to the upper. This has implications on flame front propagation and the onset of novae.
\end{abstract}

%\section{Introduction}

\noindent One of the key properties of a boundary is its `stiffness'. The `stiffness' of a convective boundary can be quantified using the bulk Richardson number, $Ri_B$, which is the ratio of the potential energy for restoration of the boundary to the kinetic energy of turbulent eddies. It is given by
\begin{gather*}
Ri_B=\frac{\Delta B\;\; L}{v^2/2},\\\\
\textrm{where} \;\;\; \Delta B(r)= \int_{r_0}^r N^2(r') dr',\\
\textrm{and}\;\;\; N^2=\frac{g\delta}{H_P}(\triangledown_{ad}-\triangledown+\frac{\phi}{\delta}\triangledown_\mu)
\end{gather*}
 
\noindent where r is the radius, r$_0$ the radial co-ordinate of the convective boundary and $L$ is the length scale that characterises turbulence and is taken to be the pressure scale height at the boundary. The final expression is the Brunt-Vaisala frequency \citep[e.g. see][]{2013sse..book.....K}, within the brackets is the familiar Ledoux criterion for convective stability. A `stiff' boundary will suppress convective boundary mixing (CBM), whereas in the opposite case a `soft' boundary will be more susceptible to CBM. Typical values of bulk Richardson numbers for `stiff' and `soft' boundaries are 10,000 and 10, respectively.

\begin{figure}
 \centering
 \includegraphics[width=\textwidth]{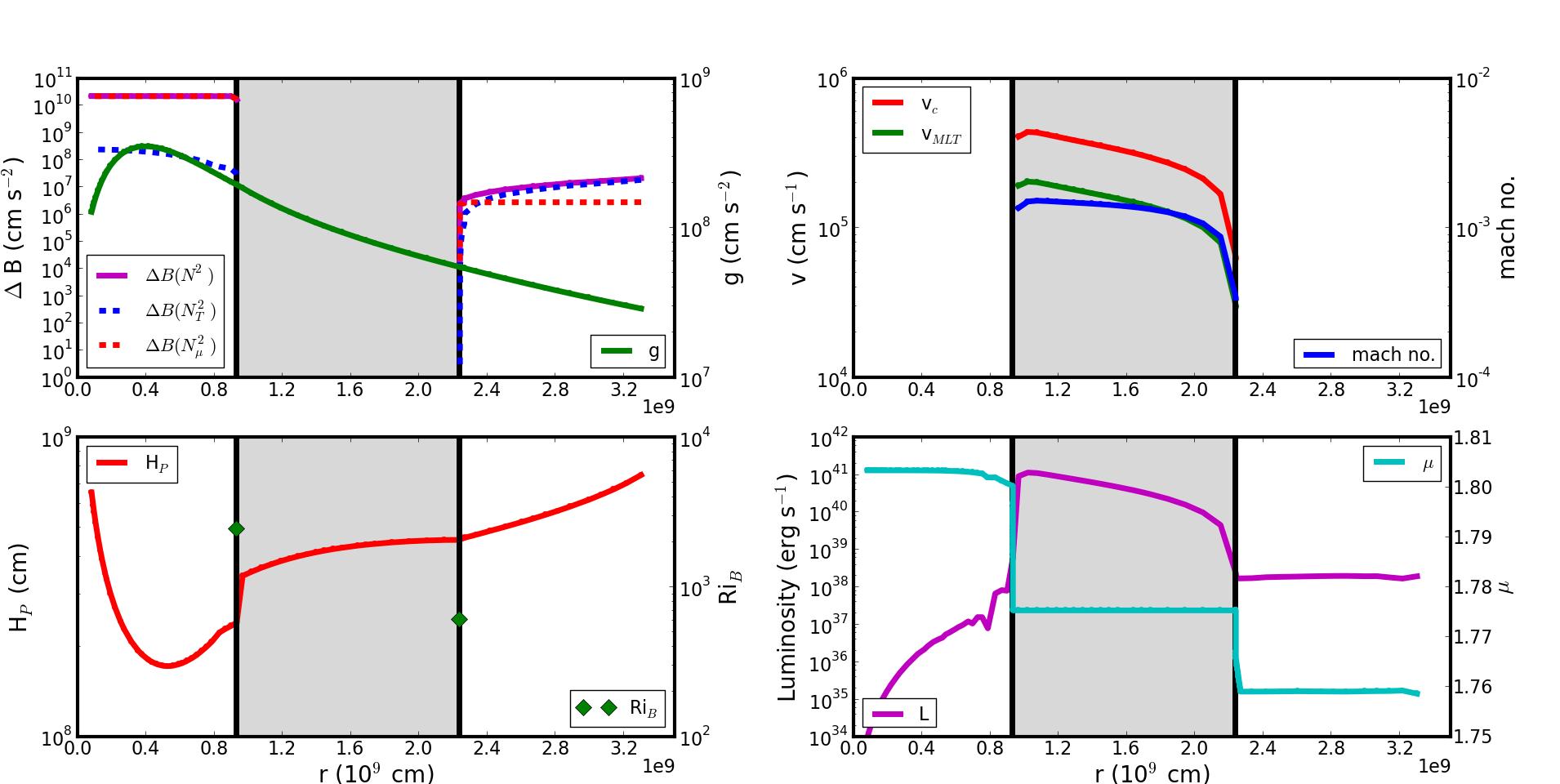}
 \caption[caption]{Structure properties of the second C shell burning region as a function of radius (31\% of the shell lifetime).\\\hspace{\textwidth} \textit{Top left} –- Buoyancy jump (magenta) and its components, thermal (blue dashed) and compositional (red dashed) and gravitational acceleration (green).\\\hspace{\textwidth} \textit{Top right} -– Convective (red), mixing length theory (green) velocities and Mach number (blue).\\\hspace{\textwidth} \textit{Bottom left} -– Pressure scale height (red) and bulk Richardson number (green diamond).\\\hspace{\textwidth} \textit{Bottom right} -– Luminosity (magenta) and mean molecular weight (cyan). Vertical black lines represent radial positions of convective boundaries and grey areas represent convective regions.}
 \label{cshell}
\end{figure}

\noindent From Fig. \ref{cshell} (bottom left panel) it can be seen that the bulk Richardson number is larger for the lower convective boundary, implying a `stiffer' boundary and suppressed convective boundary mixing (CBM). The reason for this is despite the length scale for the lower boundary been slightly smaller, the peak N value is larger for the lower boundary. The results presented here are in agreement with 3D simulations e.g. the oxygen burning shell of a 23 M$_\odot$ model by \citet{2007ApJ...667..448M}.
Suppressed CBM at lower convective boundaries has implications for other areas of astrophysics e.g. \citet{2013ApJ...762....8D} show that the onset of novae is affected by CBM, \citet{2013ApJ...772...37D} and \citet{2013ApJ...772..150J} also show that flame front propagation in S-AGB stars is affected by CBM.\\

\noindent Following a preliminary characterisation of the convective boundaries multi-D hydrodynamic simulations of convective nuclear burning shells will commence, using the code PROMPI, a parallelised version of Prometheus. Simulations are planned for the carbon and silicon burning shells of massive stars. These results will be important for the community, in understanding the advanced phases of stellar evolution, and also to produce more accurate 1D pre-supernova progenitor models.\\

\noindent The authors acknowledge support from EU-FP7-ERC-2012-St Grant 306901. R.H. acknowledges support from the World Premier International Research Center Initiative (WPI Initiative), MEXT, Japan.

\bibliographystyle{iau307}
\bibliography{references}

\end{document}